\documentclass{pnastwo}

\usepackage{epsfig}

\usepackage{amssymb,amsfonts,amsmath}

\contributor{Published in Proceedings
of the National Academy of Sciences of the United States of America}
\url{www.pnas.org/cgi/doi/10.1073/pnas.0912671107}
\copyrightyear{2010}
\issuedate{June 15, 2010}
\volume{vol. 107}
\issuenumber{no. 24}

\begin{document}

\title{Edge direction and the structure of networks}  
\date{}
\author{Jacob G. Foster\affil{1}{Complexity Science Group, Department of Physics and Astronomy, 2500 University Drive NW,
Calgary, Alberta, T2N 1N4, Canada}\affil{2}{Current affiliation: Department of Sociology, University of Chicago, 1126 East 59th St., Chicago, IL, 60615, USA.  Please direct correspondence to jgfoster@uchicago.edu}, David V. Foster\affil{1}{}, Peter Grassberger\affil{1}{}, \and Maya Paczuski\affil{1}{}}

\maketitle

\begin{article}

\begin{abstract}  Directed networks are ubiquitous and are necessary to represent complex systems with asymmetric  interactions---from food webs to the World Wide Web.  Despite the importance of edge direction for detecting local and community structure, it has been disregarded in studying a basic type of global diversity in networks: the tendency of nodes with similar numbers of edges to connect.  This tendency, called assortativity, affects crucial structural and dynamic properties of real-world networks, such as error tolerance or epidemic spreading.  Here we demonstrate that edge direction has profound effects on assortativity.  We define a set of four directed assortativity measures and assign statistical significance by comparison to randomized networks.  We apply these measures to three network classes---online/social networks, food webs, and word-adjacency networks.  Our measures (\textit{i}) reveal patterns common to each class, (\textit{ii}) separate networks that have been previously classified together, and (\textit{iii}) expose limitations of several existing theoretical models.  We reject the standard classification of directed networks as purely assortative or disassortative. Many display a class-specific mixture, likely reflecting functional or historical constraints, contingencies, and forces guiding the system's evolution.
\end{abstract}

\section{Introduction}

\dropcap{C}omplex networks reveal essential features of the structure, function, and dynamics of many complex systems \cite{BS, SIAM, BA, BN}.  While networks from diverse fields share various properties \cite{BA,SW,ER,BMod} and universal patterns \cite{BS,BA}, they also display enormous structural, functional and dynamical diversity.   A basic measure of diversity is assortativity by degree (hereafter assortativity): the tendency of nodes to link to nodes with a similar number of edges \cite{BN,newprl,newpre}).  Despite its importance, no disciplined approach to assortativity in directed networks has been proposed.  Here we present such an approach and show that measures of directed assortativity provide a number of insights into the structure of directed networks and key factors governing their evolution.

Assortativity is a standard tool in analyzing network structure \cite{BN} and has a simple interpretation. In assortative networks with symmetric interactions (i.e., undirected networks), high degree nodes, or nodes with many edges, tend to connect to other high degree nodes. Hence, assortative networks remain connected despite node removal or failure \cite{newpre}, but are hard to immunize against the spread of epidemics \cite{epi}.  In  disassortative networks, conversely, high degree nodes tend to connect to low degree nodes \cite{newprl,newpre}; these networks limit the effects of node failure because important nodes (with many edges) are isolated from each other \cite{snepprot}.  Assortativity has a convenient global measure: the Pearson correlation ($r$) between the degrees of nodes sharing an edge \cite{newprl,newpre}.  It ranges from $-1$ to $1$, with ($r>0$) in assortative networks and ($r<0$) in disassortative ones.  Earlier work proposed a simple classification of networks on the basis of assortativity, in which social networks are assortative and biological and technological networks are disassortative \cite{BN,newprl,newpre}.  Recent work suggests that this classification does not hold for undirected networks: Many online social networks are disassortative \cite{online}.  We go further, demonstrating that the simple assortative/disassortative dichotomy misses fundamental features of networks where edge direction plays a crucial role.  In fact, we show that many networks are neither purely assortative nor disassortative, but display a mixture of both tendencies.  These patterns provide a classification scheme for networks with asymmetric interactions.

In directed networks, an edge from source to target ($\rm{A} \to \rm{B}$) represents an asymmetric interaction; for example, that Web site A contains a hyperlink to Web site B, or organism A is eaten by organism B.  Edge direction is essential to evaluate and explain local structure in such networks.
For instance, motif analysis \cite{motif,mil} identifies local connection patterns that appear more frequently in the real-world network than in ensembles of randomized networks.  In this context, edge direction distinguishes functional units like feed-forward and feedback loops.   Taking edge direction into account also overturns the simple picture of the World Wide Web (WWW) as having a short average distance between all webpages \cite{diameter} in favor of a richer picture of link flow into and out of a dense inner core \cite{broder}.  More recently, attempts to identify communities in directed networks have demonstrated that ignoring edge direction misses key organizational features of community structure in networks \cite{leicht,amaral,jeong}.  Hence it is striking that assortativity in directed networks has been studied only by ignoring edge direction entirely \cite{newprl} or by measuring a subset of the four possible degree-degree correlations \cite{newpre,newdir}.  All four degree-degree correlations were addressed in the specific contexts of earthquake recurrences \cite{mayaquake} and the WWW \cite{vespignani} using the average neighbor degree, e.g. $\langle k'^{\rm{out}}\rangle_{\rm{nn}}(k^{\rm{in}})$, as a measure rather than the Pearson correlation.  However, it is easier to interpret and assign statistical significance to the Pearson correlation.  Moreover, the average neighbor degree cannot be easily used to quantify the diversity of a given network or to compare networks of various sizes, unlike the Pearson correlation.  Incorporating edge direction into familiar assortativity measures based on Pearson correlation is an essential step to better characterize, understand, and model directed networks.  Indeed, since they scale as $\cal{O}$$(E)$, where $E$ is the number of edges in the network, our directed assortativity measures can be evaluated for large networks that are beyond the reach of current motif analysis or community detection algorithms.  

Here we analyze online and social networks, food webs, and word-adjacency networks.  Classes of directed networks show common patterns across the four directed assortativity measures: $r(out,in)$; $r(in,out)$; $r(out,out)$; and $r(in,in)$. The first element in the parentheses labels the degree of the source node of the directed edge and the second labels the degree of the target node.  Thus $r(in,out)$ quantifies the tendency of nodes with high in-degree to connect to nodes with high out-degree, and so on; see Fig. 1.  

We compare the real-world network with an ensemble of randomized networks.  This comparison allows us to assign statistical significance to each measure\footnote{To our knowledge, statistical significance has been assigned to assortativity measures in only one publication \cite{flack}}.  We use that significance to define an Assortativity Significance Profile for each network.  This profile allows us to distinguish between networks grouped together by other measures; indeed, we find that online and social networks, which have similar motif structure \cite{mil}, have substantially \textit{different} assortativity profiles.  The class-specific profiles point to forces or constraints that may guide the structure, function and growth of that class \cite{mil,link,massnep}.  We also uncover limitations of several theoretical network models.  For example, neither of two plausible models of word-adjacency networks (one proposed by Milo et al. \cite{mil}, the other in this paper) can reproduce the directed assortativity profile we observe in the real-world networks.  A standard model of the WWW \cite{kra} is similarly unsuccessful.  On the other hand, the food web models \cite{mar} examined here reproduce the pattern of assortativity seen in different food webs.  Hence our measures provide useful benchmarks to test models of network formation.  

Table 1 provides descriptions and sources for all networks examined in this paper; Table 2 collects the full results including error estimates.  

\section{Results and Discussion}
 
Since nodes in directed networks have both an in-degree and an out-degree, we introduce a set of four directed assortativity measures.  Fig. 1 illustrates this set, with examples typical of assortative or disassortative networks.  Let $\alpha,\beta \in \{ in,out\}$ index the degree type, and $j_i^{\alpha}$ and $k_i^{\beta}$ be the $\alpha$- and $\beta$-degree of the source node and target node for edge $i$.  Then we define the set of assortativity measures using the Pearson correlation:

\begin{equation}
r(\alpha,\beta) = \frac{E^{-1}\sum_i[(j_i^{\alpha} - \overline{j^{\alpha}})(k_i^{\beta} - \overline{k^\beta})]}{\sigma^{\alpha}\sigma^{\beta}} 
\end{equation}
where $E$ is the number of edges in the network, $\overline{j^{\alpha}} = E^{-1} \sum_i  j_i^{\alpha}$, and $\sigma^{\alpha} = \sqrt{E^{-1}\sum_i (j_i^{\alpha} - \overline{j^{\alpha}})^2}$;  $\overline{k^{\beta}}$ and $\sigma^{\beta}$ are similarly defined.  In each correlation the edges point from the node with the $\alpha$-indexed degree to the node with the $\beta$-indexed degree (\textit{Materials and Methods}).  We assign errors by jackknife resampling \cite{newpre} and plot $2\sigma$-error bars in the figures.

To estimate statistical significance, we compare the degree-degree correlations for each real-world network to a null model.  We use as our null model the ensemble of randomized networks with the same in- and out-degree sequence (number of nodes $n(k^{\rm{in}},k^{\rm{out}})$ with in-degree $k^{\rm{in}}$ and out-degree $k^{\rm{out}}$; hereafter degree sequence) as the original network \cite{motif,mil,online,link,massnep} (\textit{Materials and Methods}).  The comparison distinguishes features accounted for by the degree sequence from those that might reflect other forces or constraints.  Our method assigns each correlation $r(\alpha,\beta)$ a statistical significance through its $Z$-score:

\begin{equation}
Z(\alpha,\beta) = \frac{r_{\rm{rw}}(\alpha,\beta) - \langle {r}_{\rm{rand}}(\alpha,\beta) \rangle}{\sigma[r_{\rm{rand}}({\alpha,\beta})]}.
\end{equation}
This quantifies the difference between the assortativity measure of the real-world network $r_{\rm{rw}}(\alpha,\beta)$ and the average assortativity for that measure in the randomized ensemble $ \langle {r}_{\rm{rand}}(\alpha,\beta) \rangle$ in units of the standard deviation $\sigma[r_{\rm{rand}}({\alpha,\beta})]$. Larger networks typically have larger $Z$ scores (see Table 2).  To compare networks of various sizes, the $Z$-scores are normalized \cite{mil} by defining an Assortativity Significance Profile (ASP), where $\textrm{ASP}(\alpha,\beta) = Z(\alpha,\beta)/(\sum_{\alpha,\beta}Z(\alpha,\beta)^2)^{1/2}$.  This quantity is directly related to the $Z$ score, and for a given network the normalization does not change the relative size of the significance measures.  To separate less significant correlations, we indicate $|Z(\alpha,\beta)|<2$ in all figures by an appropriately colored asterisk.  A positive $Z(\alpha,\beta)$ or ASP$(\alpha,\beta)$ (``$Z$ assortative") indicates that the real-world network is more assortative in that measure than expected based on the degree sequence.  A negative $Z(\alpha,\beta)$ or ASP$(\alpha,\beta)$ (``$Z$ disassortative") means that the original network is less assortative than expected.

\subsection{Online and social networks}

We first consider online and social networks.  In an online network, edges represent hyperlinks.  In the social networks considered here, edges represent positive sentiment.  Online networks are built collaboratively and share motif patterns with social networks, leading them to be grouped in the same ``superfamily" \cite{mil}.  Fig. 2a shows the ASP of the World Wide Web sample and two social networks studied in \cite{mil}.  Each network differs significantly in its ASP, showing that the ASP discriminates between networks with similar motif structure.  Fig. 2b shows the ASP of the WWW, Wikipedia \cite{(S2)}, and a network of political blogs \cite{(S3)}.  All three networks are $(out,in)$ $Z$disassortative, indicating that the small disassortative effects measured previously \cite{newpre,zla} represent substantial deviations from expected behavior.  This may reflect different growth mechanisms and/or functional constraints.  The WWW and Wikipedia are also $(in,out)$ $Z$ assortative.  This property indicates that pages with high in-degree (corresponding to ``authorities"  \cite{kle}) link to pages with high out-degree (useful pages \cite{kle}) more frequently than expected based on the degree sequence.  Pages can be both authorities and useful, and in the WWW these "multihubs" are highly interconnected; this effect creates the $(in,out)$ correlation, along with a tendency for low in-degree nodes to connect to low out-degree nodes.  All three online networks show no assortative or disassortative tendency in the $(out,out)$ or $(in,in)$ measures, consistent with previous work on the average neighbor in-degree in Wikipedia \cite{capwiki}.  The effects of $Z$-assortative or disassortative behavior can be huge, e.g. an increase of $268\%$ in the number of connections from the top $5\%$ of in-degree nodes (hereafter in-hubs) to the top $5\%$ of out-degree nodes (hereafter out-hubs) in the real-world Leadership network, compared to the randomized ensemble.  The smallest change is a $1.7\%$ decrease (blogs, in-hub to out-hub).  The $(in,out)$ effect for the WWW is substantial: an $82.3\%$ increase in connections from in-hubs to out-hubs.

Models of online network growth should reproduce the qualitative features of each online $r(\alpha,\beta)$, $Z(\alpha,\beta)$ and $\rm{ASP}(\alpha,\beta)$.  We tested a directed preferential attachment model for the WWW (\textit{Materials and Methods}) \cite{kra}. This model fails to generate any of the ASP characteristics of the WWW (Fig. 2c).  As shown in Fig. 2d, $r(in,out)$ is small in the growth model, whereas $r(in,out) = 0.2567$ is large for the WWW.  This difference arises because the growth model fails to generate many connections between multihubs orbetween low in- and low out-degree nodes.    

Thus $r(\alpha,\beta)$ and $\rm{ASP}(\alpha,\beta)$ for the three online networks cannot be attributed to the degree sequence or simple models of network growth.  The $(out,in)$ $Z$ disassortativity may reflect that hyperlinking and (more generally) information have a hierarchical structure, e.g. the existence of distinct ``high-level" topics---much as disassortativity in protein interaction networks captures the existence of weakly connected modules \cite{snepprot}.  The large $(in,out)$ assortativity and $Z$ assortativity of the WWW are especially pertinent for how users navigate the Web.  High in-degree nodes (authorities) may gain their status by aggregating links to useful pages (with high out-degree).  This pairing of trusted authorities and useful pages would provide broad access to relevant information on the Web. We find that more than half of the authorities (in-hubs) are also useful (out-hubs): Hence they may become authorities by themselves being useful.  We further find that these multihubs connect preferentially, whereas pages with low in-degree connect preferentially to pages with low out-degree.  These results are consistent with the bowtie structure revealed by a much more computationally costly analysis \cite{broder}: a densely interconnected and highly navigable core, with less trusted or useful pages clumping into small clusters or chains.  
\subsection{Food webs}

We now turn to food webs \cite{food}.  Recall that a directed edge from species A to species B means that A is eaten by B.  Food webs from diverse ecosystems display universal properties, e.g., a common form for the in- and out-degree distributions \cite{cam,dun}.  Previous work indicated that food webs are disassortative in the $(out,in)$ measure \cite{newpre}.  As shown in Fig. 3a, although $r(out,in)$ is disassortative for all food webs \cite{(S5),(S6),(S7),(S8),(S9)}, we see a wide range of values from $Z$-disassortative to $Z$-assortative in the $(out,in)$ ASP measure of Fig. 3b.  Thus, once the degree sequence is taken into account, no common pattern remains in this measure.

In contrast, food webs are both disassortative and $Z$-disassortative in the $(in,out)$ measure.  This means that organisms with many prey species are eaten by organisms with few predator species (and {\it vice versa}) more frequently than expected.  This tendency captures the structuring of ecosystems into trophic levels \cite{food}, and is consistent with an overall ``spindle" shape to the food web (fewer species in the upper and lower levels and a greater number in the middle) \cite{bas}.  The small number of species at lower trophic levels follow from the general practice of aggregating the lowest units of the food web into one or a few nodes broadly labeled ``plant," ``detritus," etc.  The consumers of these lowest units have low in-degrees and are in turn consumed by predators of low trophic level (with high out-degrees).  The food webs are assortative and $Z$ assortative in the $(out,out)$ and $(in,in)$ measures (though in the case of Ythan only slightly); since species at the same trophic level should have similar in- and out-degrees, these results may indicate that species are eating species at the same or similar trophic level---a signature of omnivory \cite{stouf}---more frequently than expected based on the degree sequence.  The effects of $Z$-assortative or disassortative behavior on linking between hubs range from a $<1\%$ increase (Little Rock, in-hub to in-hub) to a $135\%$ increase (Coachella, in-hub to in-hub).  

To identify the origin of these patterns, we built two theoretical models for each web (\textit{Materials and Methods}).  The \textit{cascade} model assigns each species a random ``niche" value and allows species to eat species of lower value with some probability \cite{mar}.  The \textit{niche} model relaxes this rigid hierarchy, permitting cannibalism and the eating of species with higher niche value \cite{mar}.  Fig. 3 c and 3d shows the $r(\alpha,\beta)$ and $\rm{ASP}(\alpha,\beta)$ for the cascade and niche models of a particular food web (St. Marks \cite{(S7)}).  The model webs shown are typical of the model, and qualitatively reproduce the pattern observed in Fig. 3 a and b.  The {\it ensemble} of niche model realizations for a given food web, however, displays large variance (see Table 3), favoring the cascade model as more predictable.  These results suggest that ordering species along a single niche dimension largely explains the observed patterns in $r(\alpha,\beta)$ and $\rm{ASP}(\alpha,\beta)$ for  food webs.  Neither model, however, typically generates the $(out,in)$ $Z$ assortativity seen in certain food webs.  

\subsection{Word-adjacency networks}
Finally, we analyze word-adjacency networks, where edges point from each word to any word that immediately follows it in a selected text \cite{mil}.  For example, ($for \to example$).  The word-adjacency networks are strongly disassortative for $r(\alpha,\beta)$; see Fig. 4a. Fig. 4b shows that they are also strongly disassortative in their ASP.  The effects on linking between high degree nodes are relatively small, ranging from a decrease of $3.8\%$ (English book, out-hub to out-hub) to a decrease of $15.8\%$ (Japanese book, out-hub to out-hub).  

The in- and the out-degree of nodes in these networks are both increasing functions of word frequency \cite{sole}; thus the correlation between a node's in-degree and out-degree is high ($r_{\rm auto} > 0.86$).  Very high frequency words generally have grammatical function but low ``semantic content" \cite{sole}.   While the large $r_{\rm auto}$ guarantees similar values for all four measures, disassortativity could result from at least two possible mechanisms.

Milo {\it et al.} propose a bipartite model (\textit{Materials and Methods}), with a few nodes of one type representing high frequency ``grammatical" words and many nodes of a second type representing low frequency content words; grammatical words must be followed by content words, and {\it vice versa} \cite{mil}.  The Bipartite  model reproduces the motif pattern of word-adjacency networks, and is thus assigned to the same superfamily in this scheme \cite{mil}.  This model generates negative values across all $r(\alpha,\beta)$, as shown in Fig. 4a , but these values are too large compared to the real network.  When compared to the rewired ensemble, however, the model reproduces the roughly equal, negative ASP$(\alpha,\beta)$ of the actual networks; see Fig. 4b.  Thus our measures do not support the classification of the Bipartite model network with the real networks.  Alternately, the observed disassortativity could result from a broad word-frequency distribution (Zipf's law \cite{sole}).  We scrambled the English text \cite{(S11)} to produce a text with identical word-frequency distribution but no grammatical structure (\textit{Materials and Methods}).  The Scrambled text  model has $r(\alpha,\beta)$ very close to the empirical values [Fig 4a], but it is $Z$-assortative across all measures [Fig 4b], unlike the real-world networks.  Neither model yields the relative magnitude of ASP$(out,in)$ and ASP$(in,out)$, suggesting that this difference results from genuine linguistic structure.

\section{Conclusions}

Our results demonstrate the fundamental importance of edge direction and the advantages  of assortativity---when properly extended---in the analysis of directed networks.  Our most basic observation is that directed networks are structurally diverse: Many directed networks are not purely assortative or disassortative, but a mixture of the two.  Our measures apply to any directed network, and we expect similar diverse but class-specific mixtures to arise in other directed networks.  By comparison with randomized ensembles, we are able to detect statistically significant features like $(in,out)$ assortativity in the WWW.  

Our measures display common patterns for classes of similar networks (see Figures 5 and 6), and can be compared to a local analogue, the Triad Significance Profile (TSP).  The TSP assigns each possible three node subgraph (motif) a normalized $Z$ score by comparing the number of appearances of the subgraph in a real-world network to the average number in a randomly rewired ensemble; classes of networks have similar TSPs \cite{mil}.  The measures $r(\alpha,\beta)$, $Z(\alpha,\beta)$ and ASP$(\alpha,\beta)$ are more computationally tractable and scalable than motif analysis; they also discriminate between networks grouped together by TSP (online/social), while confirming the motif-based classification of word-adjacency networks \cite{mil}, correctly grouping the online networks (although the political blogs only weakly), and classifying food webs together.  As illustrated by all three classes, $r(\alpha,\beta)$ and ASP$(\alpha,\beta)$ are best used together for exploring the structure of the real-world networks and testing theoretical models.  

We tested models for all three network classes.  The preferential attachment model of WWW growth \cite{kra} does not generate the observed $(in,out)$ assortativity in the WWW.  Neither the Bipartite \cite{mil} nor the Scrambled text model of word adjacency networks generates realistic patterns in both $r(\alpha,\beta)$ and ASP$(\alpha,\beta)$.  We note that creating a mixture of assortative and disassortative behavior is non-trivial.  While the WWW growth model fails to do so, both food web models \cite{mar} succeed.  We suggest that they do so by remaining close to the basic features of the phenomenon.  Our measures can be used to test models for any type of directed network and thus validate or falsify the prevailing theoretical understanding.  

The straightforward interpretation of directed assortativity leads to a variety of questions: for example, do the overabundant connections between authorities and useful pages in the WWW reflect demands of network navigation, facilitating the spread of user flow---whereas the negative $r(in,out)$ in food webs reflects the opposite tendency to concentrate energy flows at higher trophic levels?  Such questions suggest further applications of these concepts to build models better tailored to the reality of asymmetric interactions in complex networks.

\begin{materials}
\subsection{Defining the Assortativity Measures}
Newman \cite{newprl,newpre} defines $r$ in terms of the excess degree, i.e. the degree of the node minus $1$.  The correlation coefficients are exactly the same if the degree itself is used \cite{newprl}.  Identical $Z$-score results are obtained for any assortativity measure that is related to the Pearson coefficient $r(\alpha,\beta)$ by a linear transformation, e.g. the $s$ metric of Alderson and Li \cite{smetric}; thus when statistical significance is properly measured, it is sufficient to use the Pearson coefficient.  

\subsection{Constructing the Null Model}
We sample the ensemble of randomized networks with the same fixed degree sequence (FDS) \cite{motif,link,massnep} using a Monte Carlo rewiring algorithm.  The algorithm starts with a directed network with a given in- and out-degree sequence $n(k^{\rm{in}},k^{\rm{out}})$ and, by randomly swapping directed edges between nodes, samples from the FDS ensemble.  If the starting network contains self-edges, we allow them in the sampled networks; otherwise we reject such rewiring steps.  We always forbid multiple edges.  To assure random sampling, we performed $10^{5}$ edge swaps between samples for most ensembles,  $10^{6}$  for WWW and related models, and $10^{7}$ for the Wikipedia network.  Before sampling the FDS ensemble, we performed 10 times the number of intersample edge swaps on the starting network to ensure sampling of typical networks.  We assume that errors in ensemble averages are normally distributed and that after $i$ samples the difference between the mean value of an observable up to that point $\langle A \rangle_i = i^{-1}\sum_{j=1}^{i} A_j$ and the final mean $\langle A \rangle$ is less than $b\ i^{-1/2}$ in absolute value, for some constant $b$.  Plotting the difference as a function of $i^{-1/2}$ and choosing $b$ to contain approximately $90 \%$ of the data points gives an estimate of the error in the final mean, reported in Table 2 as $\sigma_r^{\rm{rand}}$.

\subsection{World Wide Web Growth Model}
The growth model for the World Wide Web is taken from \cite{kra}; we summarize it here in the original notation.  This model constructs a directed network approximating the power-law in-degree and out-degree distributions of a target real-world network, $n(k^{\rm{in}}) \propto (k^{\rm{in}})^{-\nu_{\rm{in}}}$ and $n(k^{\rm{out}}) \propto (k^{\rm{out}})^{-\nu_{\rm{out}}}$.  The model is parameterized by the number of nodes in the network, $N$; the average out-degree $\langle k^{\rm{out}}\rangle$; and the exponents, $\nu_{\rm{in}}$ and $\nu_{\rm{out}}$.  At each step, with probability $p = 1/ \langle k^{\rm{out}}\rangle$ a new node is born and attaches to an existing target node in the network, chosen with probability (depending on its in-degree $i$) $\propto A_i = i + \lambda$.  Otherwise, with probability $q = 1-p$, a directed edge is added between two existing nodes, with the source and target nodes selected with probability (depending on the out-degree of the source $j$ and in-degree of the target $i$) $\propto C(j,i) = (i+\lambda)(j+\mu)$.  Choosing $\lambda,\mu$ such that $\nu_{\rm{in}} = 2 + p \lambda$ and $\nu_{\rm{out}} = 1 + q^{-1} + \mu p q^{-1}$ generates the target exponents.  We initialize the model with two unconnected nodes and run until the network has $N$ nodes.  We eliminate multiple edges to yield a simple graph; this does not substantially alter the degree distributions or $r$ values.  For the WWW data set $\nu_{\rm{in}} = 2.32$ and $\nu_{\rm{out}} = 2.66$.  For the three model webs, the exponents are indistinguishable and are $\nu'_{\rm{in}} = 2.2 \pm 0.2$ and $\nu'_{\rm{out}} = 2.5 \pm 0.2$.  

\subsection{Cascade and Niche Models}
The food web models are taken from \cite{mar}; we summarize them here in the original notation.  Both are parameterized by the number of species in the target food web, $N$, and the connectance $C = E/N^2$, where $E$ is the number of edges.  In the cascade model, every species is assigned a random niche value chosen uniformly from $[0,1]$.  With probability $P = 2CN/(N-1)$ a species will consume a species with lower niche value.  In the niche model, every species $i$ is assigned a random niche value $n_i$ as before; the species of smallest niche value is assigned to be the ``basal species" \cite{mar}.  All other species consume every species falling within some range $r_i$.  The center of the range $c_i$ is chosen uniformly from $[0.5 r_i,n_i]$.  The range $r_i$ is chosen such that the expected connectance is that of the real-world web by setting $r_i = n_i x_i$, where $x_i$ is drawn from a beta distribution $f(x_i|1,\beta) = \beta(1-x_i)^{\beta - 1}, 0 < x_i < 1$ with expected value $E(x_i) = 1/(1+\beta) = 2C$.  Both models yield the connectance of the real-world food web, on average.  We do not check for disconnected or trophically identical species (species having identical in- and out-neighbors), as these are quite rare.  For each food web, we generated $500$ cascade model and niche model networks with $E$ within $5 \%$ of the original food web.  To identify typical networks (shown in the paper and Tables 1 and 2) we selected the model network with the smallest Euclidean distance to the ensemble average values of $r(\alpha,\beta)$.  The standard deviations in each ensemble are shown in Table 3.   

\subsection{Bipartite and Scrambled Text Models for Word-adjacency Networks}
The Bipartite model \cite{mil} assumes that there are two categories of words: a few high frequency grammatical words and many low frequency content words.  Words of the first type alternate with words of the second type, resulting in a bipartite word-adjacency network.  We build the model with $N_{\rm{gram}} = 10$ and $N_{\rm{cont}} = 1000$.  For all pairs of grammatical and content words we draw a random number $x$.  If $x< p = .06$ we put an edge from the grammatical word to the content word; if $p<x<2p$ we put an edge from the content word to the grammatical word; and if $2p< x < 2p + q$ for $q = .003$ we put an edge going each way.  The values of $p,q$ are taken from \cite{mil}.  We constructed the Scrambled Text Model by randomly scrambling the order of the words in the underlying text for one of the word-adjacency networks (English; {\it On the Origin of Species} by Charles Darwin \cite{(S11)}).  The scrambling destroys any syntactic structure, although the high frequency of articles, prepositions, etc. remains.  The assortativity across all ASP$(\alpha,\beta)$ of networks generated from the scrambled text is subtle.  The high correlation between the in- and out-degrees of a node guarantees that all values will be similar.  In the scrambled text, high frequency (high degree) words are more likely to follow one another.  But since multiple links are disallowed, rewiring, on average, destroys links between high degree nodes, making the ensemble less assortative than the Scrambled Test word-adjacency network, and making all ASP$(\alpha,\beta)$ assortative.

\end{materials}

\begin{acknowledgments}
The authors warmly thank E.A. Cartmill for her enormously helpful and detailed comments on the manuscript, K. Brown, M. Cartmill, J. Davidsen, and Seung-Woo Son for their thoughtful reading of the manuscript, the reviewers for their useful comments, and Juyong Park for an insightful discussion.  J.G.F. and P.G. acknowledge the support of iCORE.  This work was funded in part by NSERC.
\end{acknowledgments}

\section{Outline of Supporting Information}

The Supporting Information contains material of two sorts.  Figures 5 and 6 give further evidence for the classificatory ability of our measures.  Tables 1, 2, and 3: collect the basic information about all networks analyzed in the paper (1); give numerical values and error estimates for all quantities measured in the directed assortativity analysis (2); and elaborate the discussion of the cascade and niche models by showing the standard deviation for ensembles of model networks of each type (3).  We provide below an outline of the Supporting Information, briefly describing each component of the SI in order with a shortened version of the Figure or Table Legend.

\begin{itemize}
\item Figure 5: This Figure shows the similarities in ASP between several real-world networks, as measured by the dot product between their ASPs, $R_{ij} = \sum_{\alpha,\beta} \textrm{ASP}_i(\alpha,\beta)\times \textrm{ASP}_j(\alpha,\beta)$.
\item Figure 6: This Figure is constructed as in Fig. 5, but omits the $\textrm{ASP}$$(out,in)$ from the dot product.  The classes are more clearly visible in this Figure.
\item Table 1: Network properties and sources.  We show: the class of network, the number of nodes $N$, the number of edges $E$, the average out degree $\langle k_{\rm{out}} \rangle$, whether or not the network has self-edges, the Pearson correlation between the in- and out-degrees of nodes in the network $r_{\rm{auto}}$, and the source of the network.
\item Table 2: Directed assortativity results. For each network and each of the four possible pairs $(\alpha,\beta)$ we show: the Pearson correlation $r(\alpha,\beta)$, the error $\sigma_r^{\rm{rw}}$ in this quantity as estimated by jackknife \cite{newpre}, the average Pearson correlation of the random ensemble $\langle r_{\rm{rand}} \rangle $, the error of this average $\sigma_r^{\rm{rand}}$ (\textit{Materials and Methods}), $Z(\alpha,\beta)$, and ASP$(\alpha,\beta)$. 
\item Table 3: Standard deviations in food-web models.  We show the standard deviations in $r(\alpha,\beta)$ for $500$ instances per real-world network of the cascade and niche model.  Instances are constructed according to the procedure described in the \textit{Materials and Methods}, following ref. \cite{mar}; note the large standard deviations of the niche model. 
\end{itemize}

\end{article}






\begin{figure*}
\begin{center}
\includegraphics[scale=.35]{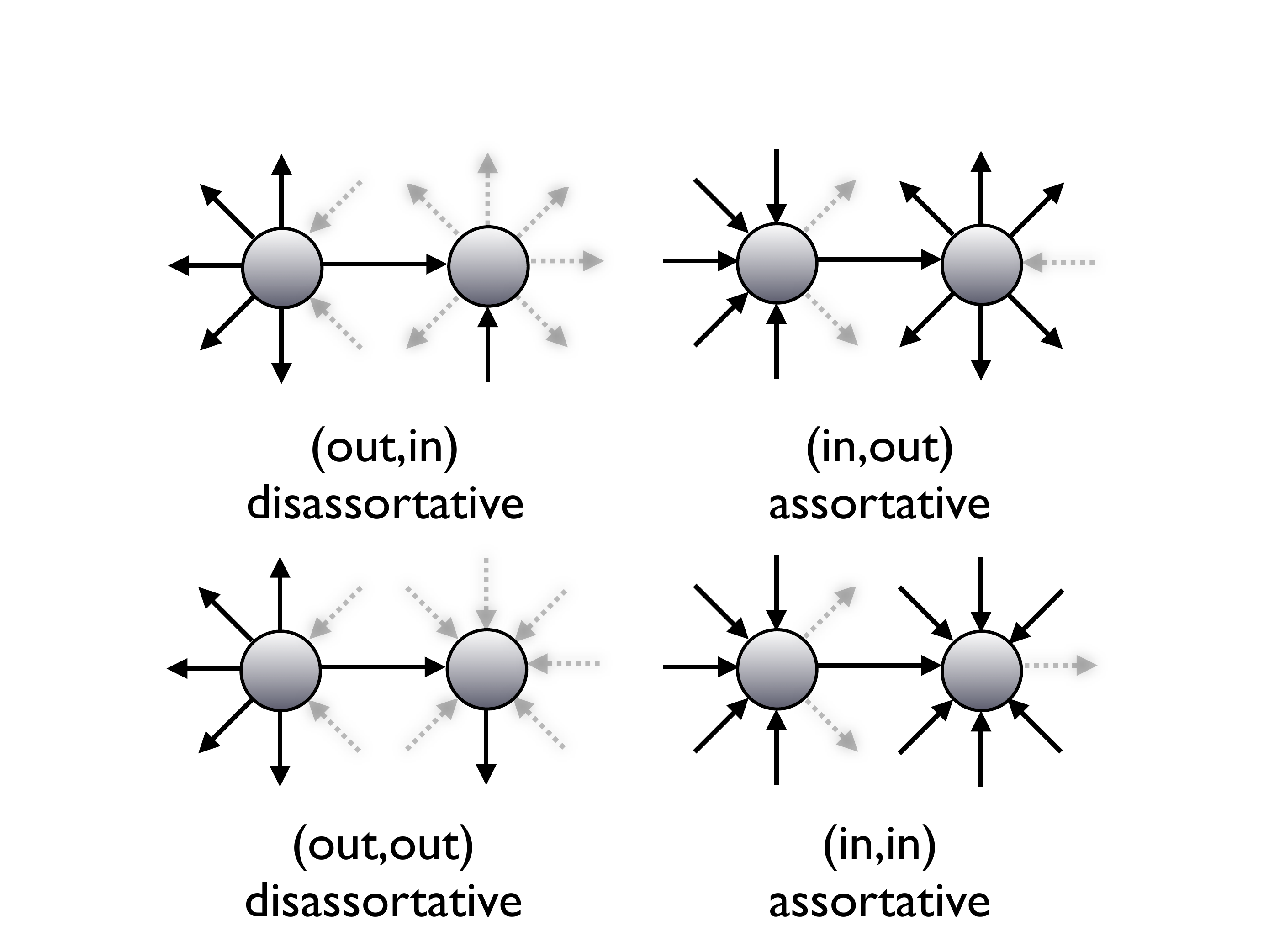}
\caption{\label{Figure 1} The four degree-degree correlations in directed networks. The fuzzy edges indicate that nodes can have any number of edges of this type, as they do not enter into the specific correlation.  For each correlation we show an example typical of assortative or disassortative networks.}
\end{center}
\end{figure*}

\begin{figure*}
\begin{center}
\includegraphics[scale=.45]{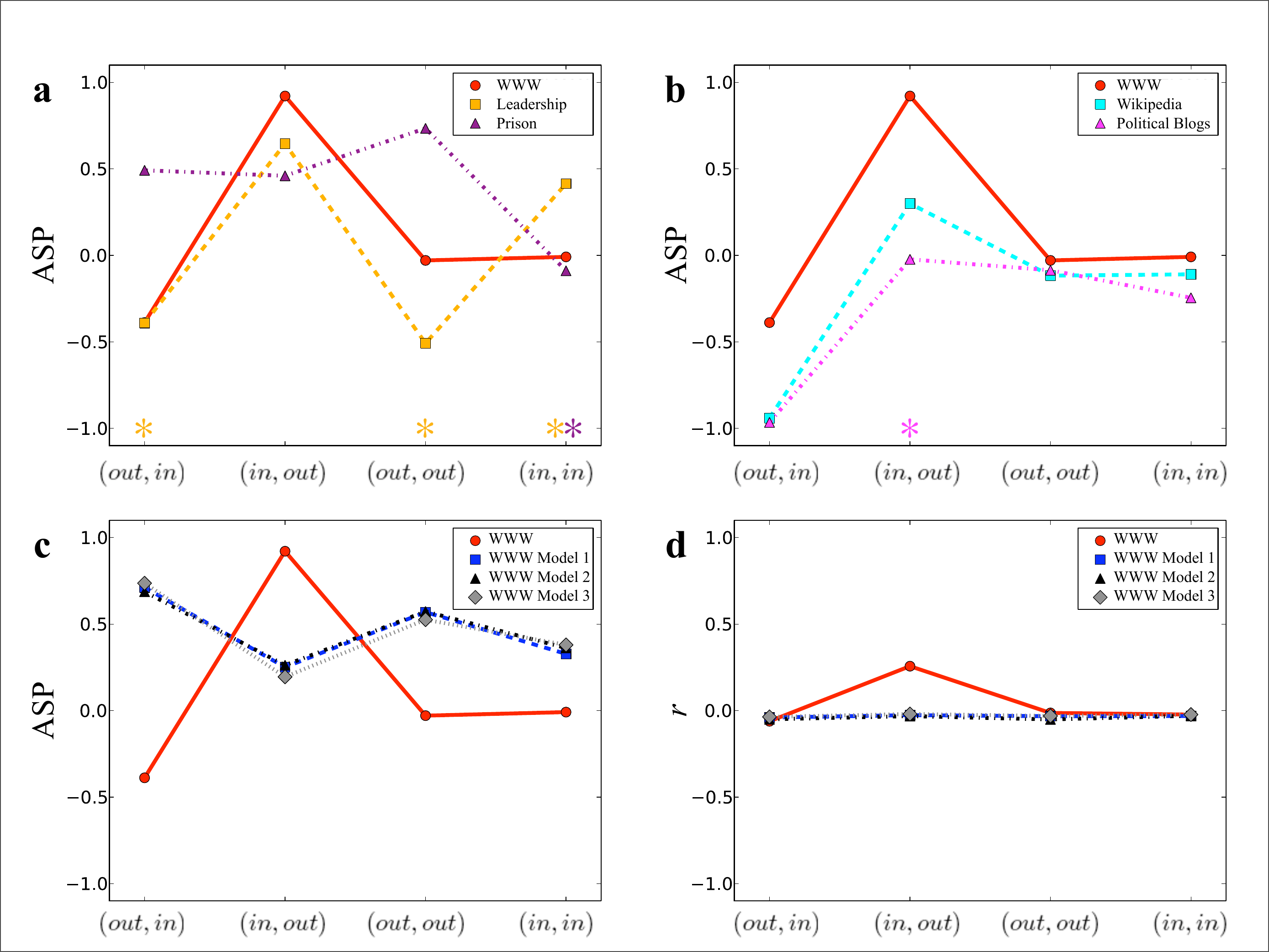}
\caption{\label{Figure 2} Online networks differ from social networks and growth models.  (\textbf{a}) The Assortativity Significance Profile (ASP) for a subset of the WWW (edges represent hyperlinks \cite{mil}) and two social networks (students in a leadership class and prisoners, edges represent positive sentiment \cite{mil}).  The three networks differ substantially, despite having similar motif patterns \cite{mil}.  In cases where $|Z|<2$, the corresponding $\rm{ASP}$ is marked with an appropriately colored asterisk.  Only Prison $(in,in)$ has $|Z| < 1$.    (\textbf{b}) The ASP for the WWW, a snapshot of Wikipedia \cite{(S2)}, and a collection of political blogs \cite{(S3)}.  All three online networks are more $(out,in)$ disassortative than expected from the degree sequence alone; the WWW and Wikipedia are significantly $(in,out)$ assortative.  The blog network has $Z(in,out) = -0.609$ and does not differ significantly from the ensemble in this measure.  All other $Z$-scores are significant. (\textbf{c} abd \textbf{d}) Three realizations of the WWW growth model \cite{kra} fail to reproduce the ASP$(\alpha,\beta)$ or $r(\alpha,\beta)$ of the WWW.  Errors in $r$, estimated via jackknife \cite{newpre}, are smaller than the symbols.}
\end{center}
\end{figure*}

\begin{figure*}
\begin{center}
\includegraphics[scale=.45]{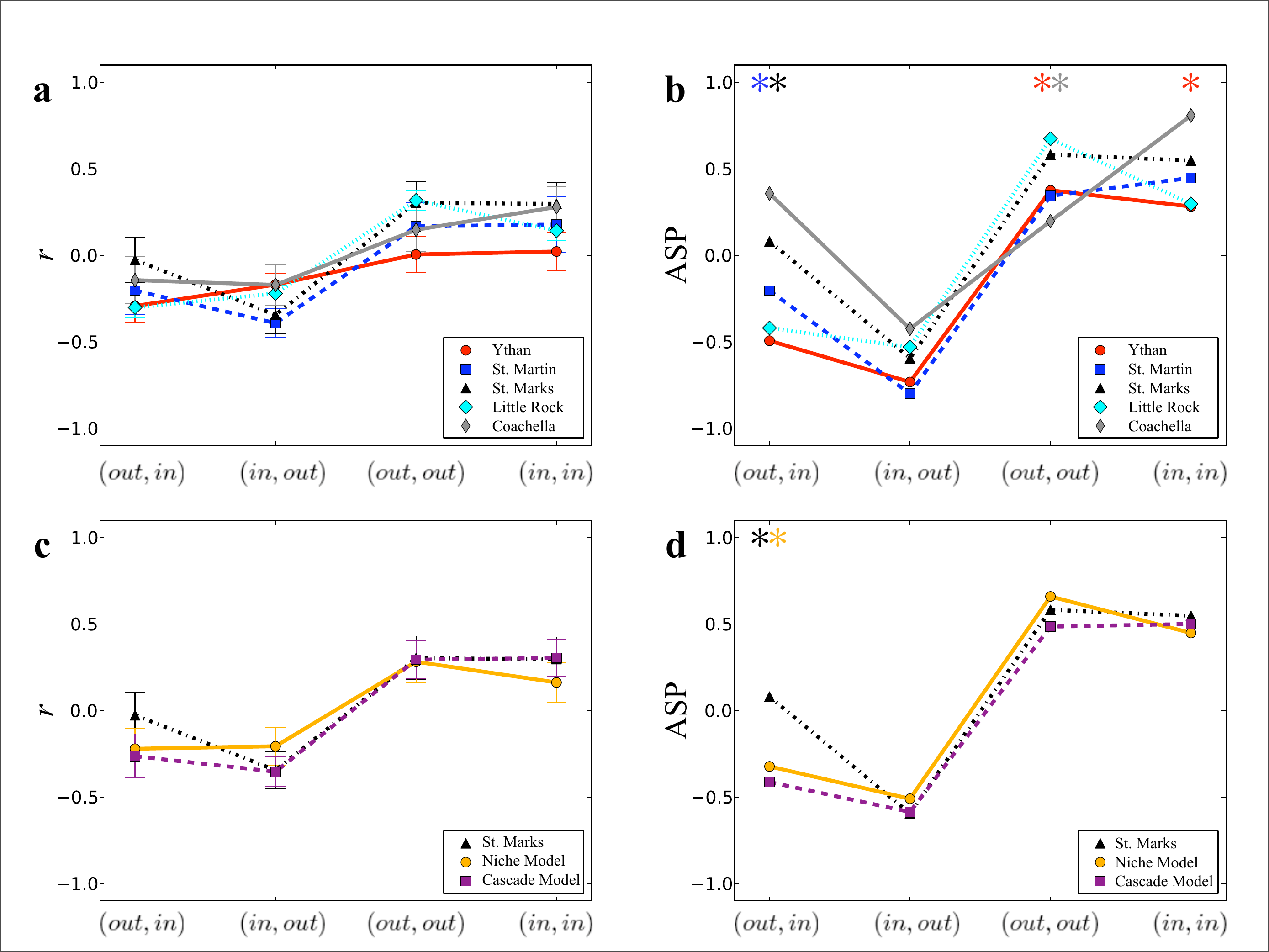}
\caption{\label{Figure 3}  Simple models largely explain directed assortativity patterns of food webs.  A directed edge from A to B indicates that A is eaten by B. (\textbf{a}) $r(\alpha,\beta)$ for food webs from several diverse ecosystems \cite{(S5),(S6),(S7),(S8),(S9)}.  Errors are estimated by jackknife \cite{newpre} and we plot $\pm 2\sigma$ error bars.  Note the common pattern: disassortative in the first two and assortative in the second two measures.  All networks save St. Marks ($(out,in)$) and Ythan ($(out,out)$, $(in,in)$) obey this pattern including errors.  (\textbf{b}) The Assortativity Significance Profile (ASP) for these food webs.  Controlling for the degree distribution highlights common $Z$-disassortative and $Z$-assortative behaviors all measures but $(out,in)$.  In cases where $|Z|<2$, the corresponding $\rm{ASP}$ is marked with an appropriately colored asterisk.  Only St. Marks $(out,in)$ has $|Z| < 1$.   (\textbf{c} and \textbf{d}) The cascade and niche models \cite{mar} reproduce most common behaviors robustly.  Errors and significance levels indicated as above. }
\end{center}
\end{figure*}

\begin{figure*}
\begin{center}
\includegraphics[scale=.45]{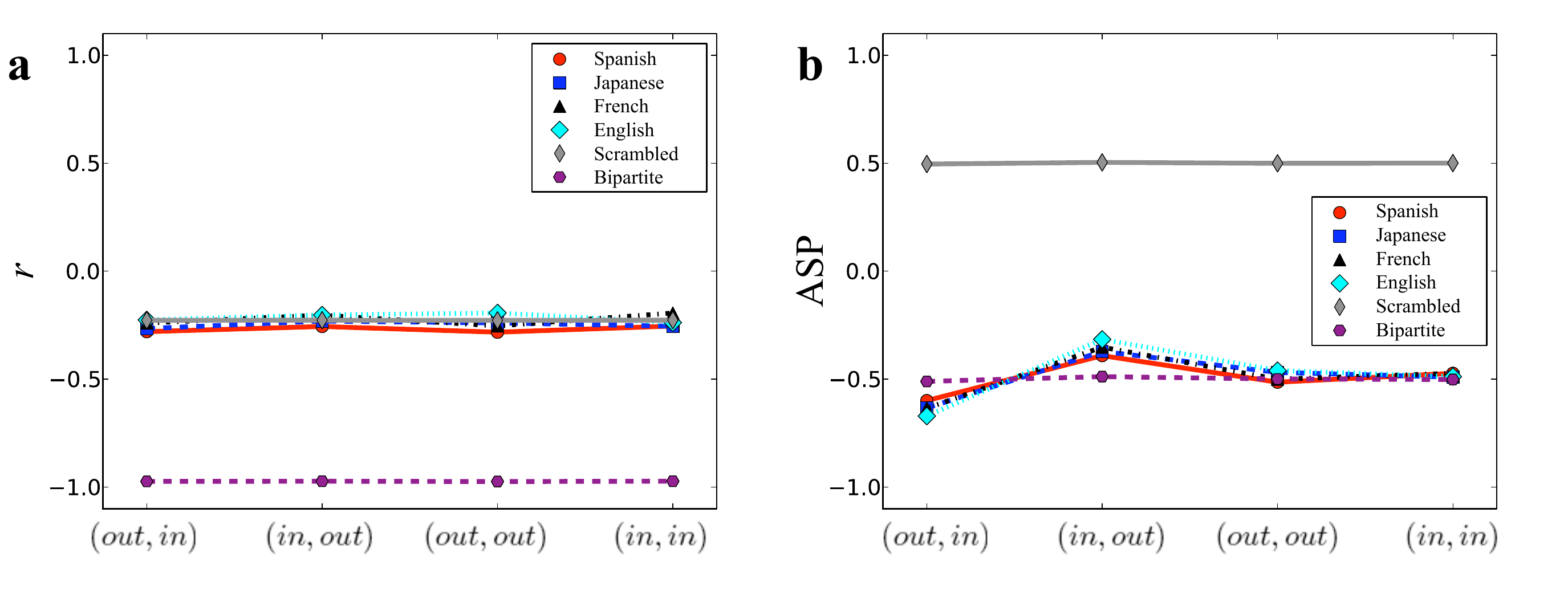}
\caption{\label{Figure 4} Simple models cannot explain directed assortativity patterns of word-adjacency networks.  A directed edge from word X to word Y indicates that X precedes Y at some point in the text under consideration.  (\textbf{a}) $r(\alpha,\beta)$ for word-adjacency networks in four languages.  The common pattern may result from grammatical structure (Bipartite model) or a broad word frequency distribution (Scrambled text model). The Bipartite model \cite{mil} overestimates the $r(\alpha,\beta)$, as shown in \textbf{a}, while the Scrambled text model \cite{(S11)} produces realistic values.  Errors in $r$ as estimated by jackknife are smaller than the symbols.  (\textbf{b}) The Assortativity Significance Profile (ASP) for the same networks.  The Bipartite model produces realistic values, while the Scrambled text model produces assortative values.   The real-world networks are remarkably similar, despite ranging in size over an order of magnitude.  All $Z$ scores are highly significant.  }
\end{center}
\end{figure*}

\begin{figure*}
\begin{center}
\includegraphics[scale=.45]{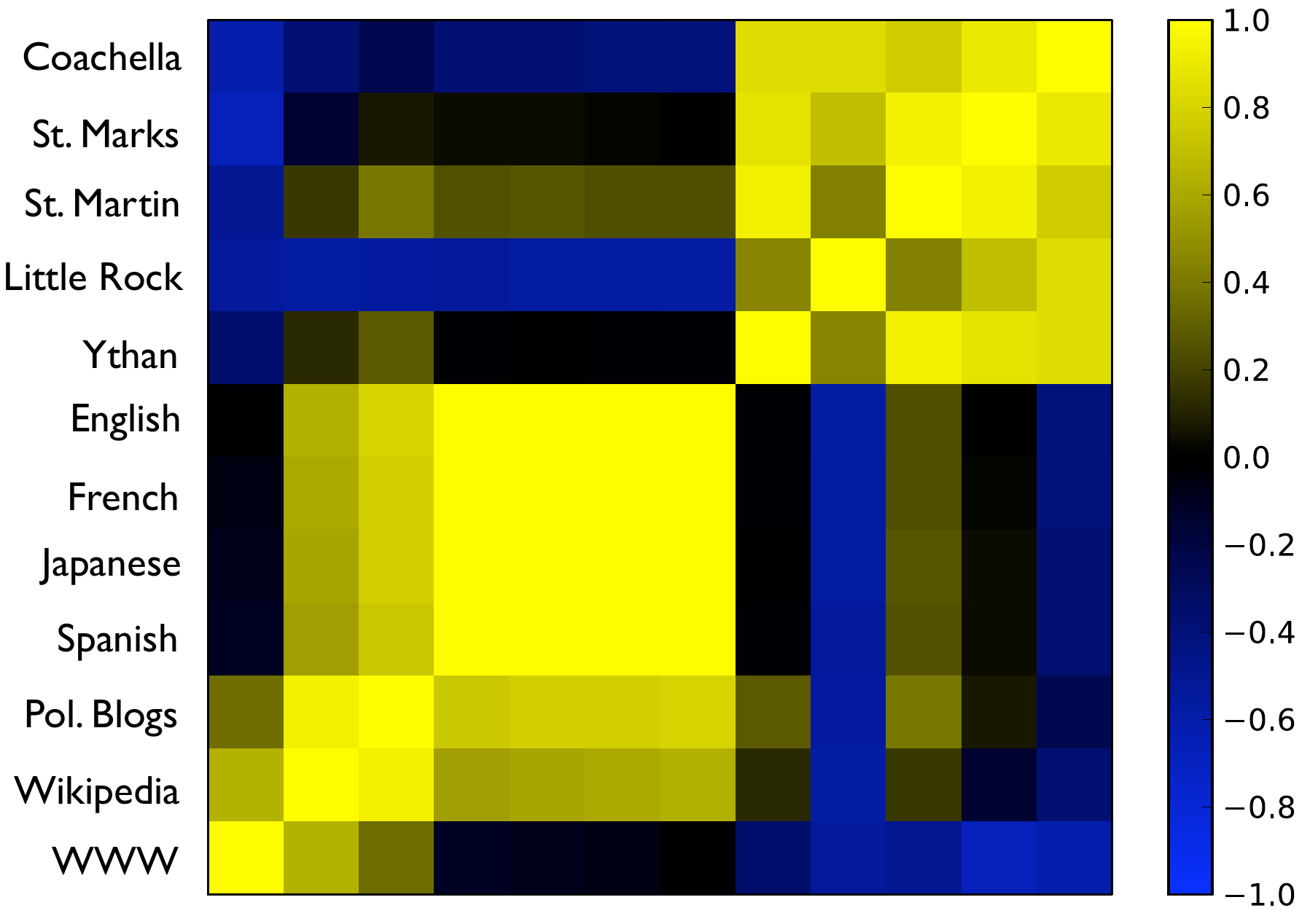}
\caption{\label{Supporting Figure 1} Similarities between several real-world networks in the ASP measure.  Each pair of real-world networks $(i,j)$ is assigned a correlation by the dot product between their ASPs, $R_{ij} = \sum_{\alpha,\beta} \textrm{ASP}_i(\alpha,\beta)\times \textrm{ASP}_j(\alpha,\beta)$.  As before, $\alpha,\beta \in \{ in,out\}$ index the degree types. Because the ASPs are normalized, $R_{ij}$ ranges from $-1$ to $1$, with $1$ indicating highly correlated ASPs.  Note that all three classes of network are clearly visible in the heat map, with some overlap between the online networks and the word-adjacency networks.  In the next Figure we identify the source of this overlap.}
\end{center}
\end{figure*}

\begin{figure*}
\begin{center}
\includegraphics[scale=.45]{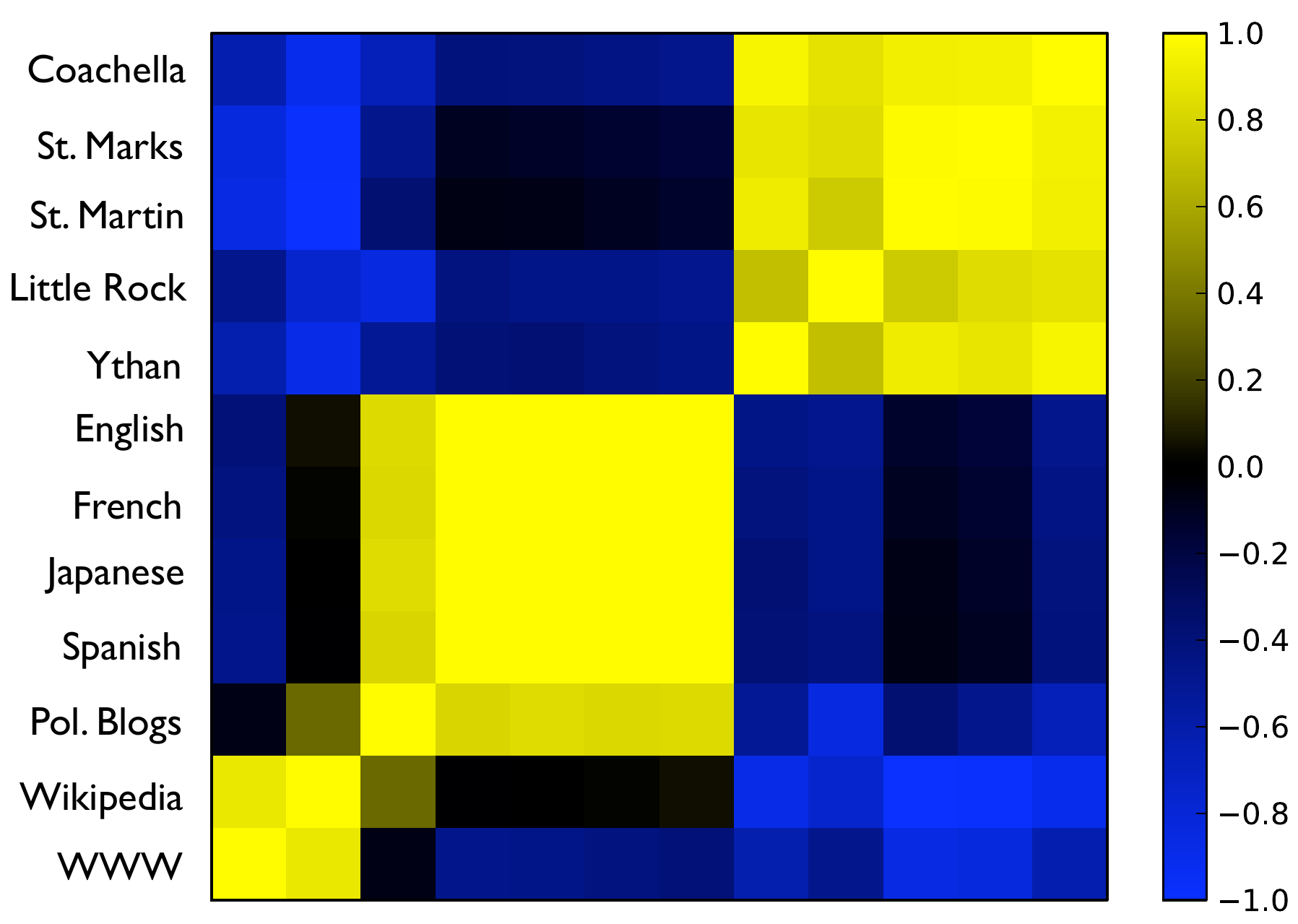}
\caption{\label{Supporting Figure 2} This is constructed as in Fig. 5, but it omits the $\textrm{ASP}$$(out,in)$ from the dot product.  The classes are much more clearly visible, which suggests that the additional measures discussed in this paper are of greater discriminatory power than the typical assortativity measure of \cite{newpre}.  Note, however, that the political blog network is not grouped with the other online networks; this is consistent with its lacking the $(in,out)$ $Z$-assortativity of the WWW and Wikipedia.}
\end{center}
\end{figure*}

\begin{table*}
\caption{Network properties and sources.  We show the class of network, the number of nodes $N$, the number of edges $E$, the average out degree $\langle k_{\rm{out}} \rangle$, whether or not the network has self-edges, the Pearson correlation between the in- and out-degrees of nodes in the network $r_{\rm{auto}}$, and the source.  Note that after reconstructing the adjacency matrix by hand from refs. \cite{(S5), (S6),(S7),(S8),(S9)}, we performed a trophic aggregation on all food webs, meaning that if two species had identical interactions, we combined them into one node.  Further, all parasites were removed from the Ythan food web.\newline }
\begin{tabular}{|c| | |c|c|c|c|c|c|c|c|}
\hline
Network & Class & $N$ &$E$ &$\langle k_{\rm{out}} \rangle $& Self-edges &  $r_{\rm{auto}}$ & Source \\
\hline \hline \hline
Leadership & social & 32 & 96 & 3.000 & No &  0.053 & \cite{mil}  \\
\hline
Prison & social & 67 & 182 & 2.716 & No &  0.201 & \cite{mil}  \\
\hline
WWW & online & 325729 & 1497135 & 4.596 & Yes &  0.211 & \cite{mil}  \\
\hline
Wikipedia & online & 1598583 & 19753078 & 12.357 & Yes &  0.203 & \cite{(S2)}  \\
\hline
Pol. Blogs & online & 1224 & 19090 & 15.597 & Yes &  0.377 & \cite{(S3)}  \\
\hline
WWW Model 1 & online & 325729 & 1446887 & 4.442 & Yes &  0.526 & \cite{kra}  \\
\hline
WWW Model 2 & online & 325729 & 1448691 & 4.448 & Yes &  0.565 & \cite{kra}  \\
\hline
WWW Model 3 & online & 325729 & 1428052 & 4.384 & Yes &  0.391 & \cite{kra}  \\
\hline
Coachella & food web & 29 & 262 & 9.034 & Yes &  -0.361 & \cite{(S5)}  \\
\hline
Little Rock & food web & 95 & 1080 & 11.368 & Yes &  -0.242 & \cite{(S6)}  \\
\hline
St. Marks & food web & 48 & 221 & 4.604 & Yes &  -0.227 & \cite{(S7)}  \\
\hline
St. Martin & food web & 42 & 205 & 4.881 & No &  -0.368 & \cite{(S8)}  \\
\hline
Ythan & food web & 82 & 395 & 4.817 & Yes &  -0.055 & \cite{(S9)}  \\
\hline
Coachella Niche & food web & 29 & 259 & 8.931 & Yes &  -0.408 & \cite{mar}  \\
\hline
Little Rock Niche & food web & 95 & 1056 & 11.116 & Yes &  -0.284 & \cite{mar}  \\
\hline
St. Marks Niche & food web & 48 & 216 & 4.500 & Yes &  -0.258 & \cite{mar}  \\
\hline
St. Martin Niche & food web & 41 & 208 & 5.073 & No &  -0.398 & \cite{mar}  \\
\hline
Ythan Niche & food web & 82 & 386 & 4.707 & Yes &  -0.389 & \cite{mar}  \\
\hline
Coachella Cascade & food web & 29 & 267 & 9.207 & No &  -0.907 & \cite{mar} \\
\hline
Little Rock Cascade & food web & 95 & 1098 & 11.558 & No &  -0.859 &\cite{mar} \\
\hline
St. Marks Cascade & food web & 48 & 223 & 4.646 & No &  -0.793 & \cite{mar}  \\
\hline
St. Martin Cascade & food web & 42 & 205 & 4.881 & No &  -0.662 & \cite{mar}  \\
\hline
Ythan Cascade & food web & 82 & 384 & 4.683 & No &  -0.702 & \cite{mar}  \\
\hline
Spanish & word adj. & 11586 & 45129 & 3.895 & No &  0.913 & \cite{mil}  \\
\hline
Japanese & word adj. & 2704 & 8300 & 3.070 & No &  0.927 & \cite{mil}  \\
\hline
French & word adj. & 8325 & 24295 & 2.918 & No &  0.905 & \cite{mil}  \\
\hline
English & word adj. & 8525 & 74921 & 8.788 & Yes &  0.876 & \cite{(S11)}  \\
\hline
Scrambled & word adj. & 8525 & 118161 & 13.861 & Yes &  0.999 & \cite{(S11)}  \\
\hline
Bipartite & word adj. & 746 & 1290 & 1.729 & No &  0.968 & \cite{mil}  \\
\hline
\end{tabular}
\end{table*}

\begin{table*}

\caption{Directed assortativity results. For each network and each of the four possible pairs $(\alpha,\beta)$ we show the Pearson correlation $r(\alpha,\beta)$, the error $\sigma_r^{\rm{rw}}$ in this quantity as estimated by jackknife \cite{newpre}, the average Pearson correlation of the random ensemble $\langle r_{\rm{rand}} \rangle $, the error of this average $\sigma_r^{\rm{rand}}$ (\textit{Materials and Methods}), $Z(\alpha,\beta)$, and ASP$(\alpha,\beta)$.  \newline}

\begin{tabular}{|c|c| | c|c|c|c|c|c|c|}
\hline
Network & $(\alpha,\beta)$ & $r(\alpha,\beta)$ &$\sigma_r^{\rm{rw}}$ &$\langle r_{\rm{rand}} \rangle $&$\sigma_r^{\rm{rand}}$ &  $Z(\alpha,\beta)$ & ASP$(\alpha,\beta)$ \\
\hline \hline \hline
Leadership & $(out,in)$ & -0.157 & 0.123 & -0.030 & 0.0015 &  -1.419 & -0.391 \\
		   & $(in,out)$ & 0.214 & 0.107 & -0.015 & 0.0014 &  2.344 & 0.646 \\
		   & $(out,out)$ & -0.199 & 0.010 & -0.036 & 0.0013 &  -1.844 & -0.508  \\
		   & $(in,in)$ & -0.083 & 0.089 & -0.045 & 0.0013 &  1.504 & 0.415  \\
		   \hline
Prison & $(out,in)$ & 0.129 & 0.072 & -0.023 & 0.0010 &  2.152 & 0.492 \\
		   & $(in,out)$ & 0.134 & 0.067 & -0.012 & 0.0016 &  2.013 & 0.460 \\
		   & $(out,out)$ & 0.206 & 0.073 & -0.021 & 0.0016 &  3.214 & 0.734  \\
		   & $(in,in)$ & -0.053 & 0.070 & -0.027 &  0.0016 &  -0.390 & -0.089  \\
		   \hline
WWW & $(out,in)$ & -0.062 & 0.0001 & -0.039 & $3.0 \times 10^{-6}$ &  -144.927 & -0.388 \\
		   & $(in,out)$ & 0.257 & 0.0002 & 0.000 & $1.8 \times 10^{-5}$ &  343.609 & 0.921 \\
		   & $(out,out)$ & -0.014& 0.0001 & -0.007 & $1.7 \times 10^{-5}$ &  -10.861 &-0.029  \\
		   & $(in,in)$ & -0.023 & 0.0001 & -0.021 & $1.5 \times 10^{-5}$ &  -3.258 & -0.009  \\  
\hline
Wikipedia & $(out,in)$ & -0.070 & 0.0002 & -0.037 & $3.8 \times 10^{-6}$&  -392.737 & -0.941 \\
		   & $(in,out)$ & 0.017 & 0.0028 & -0.005 & $2.8 \times 10^{-5}$ &  125.057 & 0.299 \\
		   & $(out,out)$ & -0.032 & 0.0006 & -0.024 & $3.0 \times 10^{-5}$ &  -48.970 & -0.117  \\
		   & $(in,in)$ & -0.014 & 0.0008 & -0.009 & $6.0 \times 10^{-6}$ &  -45.744 & -0.110 \\  
\hline
Pol. Blogs & $(out,in)$ & -0.230 & 0.005 & -0.133 & $4.5 \times 10^{-5}$ &  -25.689 & -0.965 \\
		   & $(in,out)$ & -0.023 & 0.006 & -0.020 & $5.8 \times 10^{-5}$ &  -0.609 & -0.023 \\
		   & $(out,out)$ & -0.0515 & 0.006 & -0.041 & $6.5 \times 10^{-5}$ &  -2.285 & -0.086  \\
		   & $(in,in)$ & -0.094 & 0.006 & -0.064 & $7.6 \times 10^{-5}$ &  -6.522 & -0.245  \\  
\hline
WWW Model 1& $(out,in)$ & -0.040 & 0.0001 & -0.043 & $4.5 \times 10^{-7}$ &  77.186 & 0.711 \\
		   & $(in,out)$ & -0.026 & 0.0003 & -0.029 & $5.0 \times 10^{-6}$ &  27.230 & 0.251 \\
		   & $(out,out)$ & -0.033 & 0.0002 & -0.037 & $8.0 \times 10^{-6}$ &  61.734 & 0.570  \\
		   & $(in,in)$ & -0.031 & 0.0002 & -0.033 & $7.5 \times 10^{-7}$ &  35.574 & 0.328  \\  
\hline
WWW Model 2& $(out,in)$ & -0.050 & 0.0002 & -0.054 & $6.5 \times 10^{-7}$ &  77.496 & 0.687 \\
		   & $(in,out)$ & -0.032 & 0.0003 & -0.036 & $4.5 \times 10^{-6}$ &  29.586 & 0.262 \\
		   & $(out,out)$ & -0.051 & 0.0003 & -0.060 & $1.8 \times 10^{-5}$ &  64.594 & 0.573  \\
		   & $(in,in)$ & -0.030 & 0.0002 & -0.031 & $6.7 \times 10^{-7}$ &  40.795 & 0.362 \\  
\hline
WWW Model 3& $(out,in)$ & -0.036 & 0.0001 & -0.037 & $1.9 \times 10^{-7}$ &  73.870 & 0.736 \\
		   & $(in,out)$ & -0.020 & 0.0003 & -0.021 & $1.5 \times 10^{-6}$ &  19.573 & 0.195 \\
		   & $(out,out)$ & -0.031 & 0.0002 & -0.033 & $4.5 \times 10^{-6}$ &  52.737 & 0.525  \\
		   & $(in,in)$ & -0.023 & 0.0001 & -0.024 &  $1.4 \times 10^{-7}$  &  38.111 & 0.380  \\   
		   \hline
\end{tabular}
\end{table*}

\begin{table*}
\begin{tabular}{|c|c| | c|c|c|c|c|c|c|}
\hline
Network & $(\alpha,\beta)$ & $r(\alpha,\beta)$ &$\sigma_r^{\rm{rw}}$ &$\langle r_{\rm{rand}} \rangle $&$\sigma_r^{\rm{rand}}$ &  $Z(\alpha,\beta)$ & ASP$(\alpha,\beta)$ \\
\hline \hline \hline
Coachella& $(out,in)$ & -0.143 & 0.068 & -0.229 &$5.3 \times 10^{-4}$&  2.642 & 0.357 \\
		   & $(in,out)$ & -0.170 & 0.059 & -0.037 & $4.7 \times 10^{-4}$ &  -3.134 & -0.424 \\
		   & $(out,out)$ & 0.148 & 0.063 &0.096 & $4.2 \times 10^{-4}$ &  1.459 & 0.197  \\
		   & $(in,in)$ & 0.280 & 0.058 & 0.055 & $6.2 \times 10^{-4}$ &  5.971 & 0.808  \\   
		   \hline

Little Rock& $(out,in)$ & -0.301& 0.030 & -0.197 & $2.3 \times 10^{-4}$ &  -5.902 & -0.420 \\
		   & $(in,out)$ & -0.221 & 0.025 & -0.029 & $2.6 \times 10^{-4}$ &  -7.464 & -0.531 \\
		   & $(out,out)$ & 0.317 & 0.029 &0.098 & $2.6 \times 10^{-4}$ &  9.476 & 0.672  \\
		   & $(in,in)$ & 0.142 & 0.029 & 0.049 & $4.3 \times 10^{-4}$ &  4.181 & 0.297  \\   
		   \hline
St. Marks& $(out,in)$ & -0.027& 0.065 & -0.069 & $5.7 \times 10^{-4}$ &  0.735 & 0.081 \\
		   & $(in,out)$ & -0.344 & 0.054 & -0.011 & $6.6 \times 10^{-4}$  &  -5.390 & -0.595 \\
		   & $(out,out)$ & 0.302 & 0.061 &-0.010 & $6.7 \times 10^{-4}$  &  5.280 & 0.583  \\
		   & $(in,in)$ & 0.298 & 0.061 & 0.004 & 0.00115 &  4.964 & 0.548\\   
		   \hline
St. Martin& $(out,in)$ & -0.204& 0.068 & -0.127 & $7.2 \times 10^{-4}$ &  -1.476 &-0.204 \\
		   & $(in,out)$ & -0.392 & 0.042 & -0.020 &  $9.2 \times 10^{-4}$ &  -5.790 & -0.800 \\
		   & $(out,out)$ & 0.168 & 0.069 &0.017 & $9.2 \times 10^{-4}$ &  2.492 & 0.344 \\
		   & $(in,in)$ & 0.178 & 0.081 & 0.014 & $8.5 \times 10^{-4}$ &  3.244 & 0.448 \\   
		   \hline
Ythan& 	       $(out,in)$ & -0.179 & 0.047 & -0.238 & $3.0 \times 10^{-4}$ &  -2.308 & -0.493 \\
		   & $(in,out)$ & -0.338 & 0.033 & -0.014 & $6.1 \times 10^{-4}$ &  -3.424 & -0.732 \\
		   & $(out,out)$ & 0.348 & 0.052 &-0.062 & $6.1 \times 10^{-4}$ &  1.759 & 0.376  \\
		   & $(in,in)$ & 0.288 & 0.056 & -0.017& $2.9 \times 10^{-4}$ &  1.321 & 0.282  \\   
		   \hline

Coachella Niche& $(out,in)$ & -0.143 & 0.063 & -0.195 & $7.4 \times 10^{-4}$ &  0.505 & 0.045 \\
		   & $(in,out)$ & -0.170 & 0.043 & -0.020 & $5.6 \times 10^{-4}$ &  -6.383 & -0.573 \\
		   & $(out,out)$ & 0.148 & 0.049 &0.085 & $5.4 \times 10^{-4}$ &  5.866 & 0.527  \\
		   & $(in,in)$ & 0.280 & 0.061 & 0.031 & $6.6 \times 10^{-4}$ &  6.969 & 0.626  \\   
		   \hline

Little Rock Niche& $(out,in)$ & -0.206& 0.030 & -0.073 & $4.2 \times 10^{-4}$ &  -5.197 & -0.288 \\
		   & $(in,out)$ & -0.263 & 0.027 & -0.006 & $3.4 \times 10^{-4}$ &  -9.467 & -0.524 \\
		   & $(out,out)$ & 0.337 & 0.027 &0.013 & $3.3 \times 10^{-4}$ & 12.131  & 0.671  \\
		   & $(in,in)$ & 0.198 & 0.030 & 0.001 & $3.3 \times 10^{-4}$ &  7.914 & 0.438  \\   
		   \hline
St. Marks Niche& $(out,in)$ & -0.221& 0.059 & -0.113 & 0.00124 &  -1.964 & -0.323 \\
		   & $(in,out)$ & -0.206 & 0.055 & -0.013 & 0.00105 &  -3.099 & -0.509 \\
		   & $(out,out)$ & 0.282 & 0.061 &0.046 & $8.6 \times 10^{-4}$&  4.014 & 0.660 \\
		   & $(in,in)$ & 0.163 & 0.066 & 0.004 &$8.5 \times 10^{-4}$ &  2.730 & 0.449\\   
		   \hline
St. Martin Niche& $(out,in)$ & -0.230& 0.066 & -0.181 & $4.4 \times 10^{-4}$ &  -1.230 &-0.225 \\
		   & $(in,out)$ & -0.221 & 0.043 & -0.038 & $5.6 \times 10^{-4}$ &  -2.926 & -0.536 \\
		   & $(out,out)$ & 0.312 & 0.062 &0.083 & $5.3 \times 10^{-4}$ &  3.911 & 0.716\\
		   & $(in,in)$ & 0.182 & 0.081 & 0.067 & $9.1 \times 10^{-4}$ &  2.106 & 0.386\\   
		   \hline
Ythan Niche& 	       $(out,in)$ & -0.193 & 0.058 & -0.074 & $5.7 \times 10^{-4}$ &  -2.443 & -0.324 \\
		   & $(in,out)$ & -0.243 & 0.037 & -0.018 & $5.2 \times 10^{-4}$ &  -4.728 & -0.616 \\
		   & $(out,out)$ & 0.252 & 0.046 & 0.043 & $5.2 \times 10^{-4}$ &  4.414 & 0.585  \\
		   & $(in,in)$ & 0.158 & 0.060 & 0.020& $5.7 \times 10^{-4}$ &  3.034 & 0.402  \\   
		   \hline
\end{tabular}
\end{table*}

\begin{table*}
\begin{tabular}{|c|c| | c|c|c|c|c|c|c|}
\hline
Network & $(\alpha,\beta)$ & $r(\alpha,\beta)$ &$\sigma_r^{\rm{rw}}$ &$\langle r_{\rm{rand}} \rangle $&$\sigma_r^{\rm{rand}}$ &  $Z(\alpha,\beta)$ & ASP$(\alpha,\beta)$ \\
\hline \hline \hline
Coachella Cascade& $(out,in)$ & -0.415 & 0.050 & -0.229 & $4.7 \times 10^{-4}$ & -5.713 & -0.453 \\
		   & $(in,out)$ & -0.458 & 0.038 & -0.037 & $2.1 \times 10^{-4}$ &  -6.891 & -0.547 \\
		   & $(out,out)$ & 0.436 & 0.048 &0.096 & $3.2 \times 10^{-4}$ &  6.383 & 0.506  \\
		   & $(in,in)$ & 0.433 & 0.043 & 0.055 & $3.8 \times 10^{-4}$ &  6.173 & 0.490  \\   
		   \hline
Little Rock Cascade& $(out,in)$ & -0.363& 0.027 & -0.051 & $4.1 \times 10^{-4}$ &  -11.977 & -0.465 \\
		   & $(in,out)$ & -0.417 & 0.020 & -0.034 & $2.1 \times 10^{-4}$ &  -13.735 & -0.533 \\
		   & $(out,out)$ & 0.389 & 0.025 &0.041 & $2.0 \times 10^{-4}$ &  12.756 & 0.495  \\
		   & $(in,in)$ & 0.391 & 0.024 & 0.039 & $3.8 \times 10^{-4}$ &  13.033 & 0.506 \\   
		   \hline
St. Marks Cascade& $(out,in)$ & -0.264& 0.062 & -0.040 & $9.2 \times 10^{-4}$ &  -3.627 & -0.413 \\
		   & $(in,out)$ & -0.353 & 0.043 & -0.020 & $6.7 \times 10^{-4}$ &  -5.146 & -0.586 \\
		   & $(out,out)$ & 0.294 & 0.055 &0.025 & $6.7 \times 10^{-4}$&  4.260 & 0.485 \\
		   & $(in,in)$ & 0.305 & 0.053 & 0.024 &$ 7.5 \times 10^{-4}$ &  4.398 & 0.501\\   
		   \hline
St. Martin Cascade& $(out,in)$ & -0.289& 0.066 & -0.056 & $9.2 \times 10^{-4}$ &  -3.821 &-0.424 \\
		   & $(in,out)$ & -0.371 & 0.056 & -0.021 & $7.7 \times 10^{-4}$ &  -5.293 & -0.587 \\
		   & $(out,out)$ & 0.310 & 0.055 &0.022 & $7.7 \times 10^{-4}$ & 4.536 & 0.503 \\
		   & $(in,in)$ & 0.297 & 0.065 & 0.026 & 0.00145 & 4.265 & 0.473 \\   
		   \hline
Ythan Cascade& 	       $(out,in)$ & -0.257 & 0.046 & -0.023 & $8.7 \times 10^{-4}$ &  -4.873 & -0.431 \\
		   & $(in,out)$ & -0.346 & 0.041 & -0.011 & $6.5 \times 10^{-4}$ &  -6.703 & -0.592 \\
		   & $(out,out)$ & 0.275 & 0.044 &0.012 & $6.5 \times 10^{-4}$ &  5.401 & 0.477 \\
		   & $(in,in)$ & 0.283 & 0.045 & 0.010& $9.3 \times 10^{-4}$ &  5.495 & 0.486  \\   
		   \hline

Spanish& $(out,in)$ & -0.280 & 0.002 & -0.269 & $3.8 \times 10^{-6}$ &  -75.777 & -0.599 \\
		   & $(in,out)$ & -0.256 & 0.002 & -0.246 & $4.7 \times 10^{-6}$ &  -49.451 & -0.391 \\
		   & $(out,out)$ & -0.282 & 0.002 &-0.269 & $2.4 \times 10^{-5}$ &  -65.006 & -0.514  \\
		   & $(in,in)$ & -0.254 & 0.002 & -0.246 & $3.8 \times 10^{-6}$ &  -59.801 & -0.473 \\   
		   \hline

Japanese& $(out,in)$ & -0.266 & 0.004 & -0.230 & $1.9 \times 10^{-5}$ &  -29.772 & -0.634 \\
		   & $(in,out)$ & -0.231 & 0.004 & -0.208 & $2.8 \times 10^{-5}$ &  -17.468 & -0.372 \\
		   & $(out,out)$ & -0.240 & 0.004 &-0.213 & $2.9 \times 10^{-5}$ &  -22.025 & -0.469  \\
		   & $(in,in)$ & -0.255 & 0.004 & -0.224 & $3.0 \times 10^{-5}$ &  -23.062 & -0.491 \\   
		   \hline
French& $(out,in)$ & -0.240 & 0.002 & -0.210 & $6.2 \times 10^{-6}$ &  -75.777 & -0.599 \\
		   & $(in,out)$ & -0.204 & 0.002 & -0.183 & $1.3 \times 10^{-5}$ &  -49.451 & -0.391 \\
		   & $(out,out)$ & -0.253 & 0.002 &-0.220 & $2.8 \times 10^{-5}$ &  -65.006 & -0.514  \\
		   & $(in,in)$ & -0.194 & 0.002 & -0.174 & $4.8 \times 10^{-6}$ &  -59.801 & -0.473 \\   
		   \hline

English& $(out,in)$ & -0.226 & 0.001 & -0.214 & $3.3 \times 10^{-6}$ &  -69.192 & -0.671 \\
		   & $(in,out)$ & -0.203 & 0.001 & -0.195 & $5.7 \times 10^{-6}$ &  -32.554 & -0.316 \\
		   & $(out,out)$ & -0.193 & 0.001 &-0.185 & $9.7 \times 10^{-6}$ &  -47.468 & -0.460  \\
		   & $(in,in)$ & -0.238 & 0.001 & -0.227 & $3.9 \times 10^{-6}$ &  -50.332 & -0.488 \\   
		   \hline

Scrambled& $(out,in)$ & -0.227 & 0.001 & -0.235 & $4.3 \times 10^{-6}$ &  43.805 & 0.496 \\
		   & $(in,out)$ & -0.227 & 0.001 & -0.235 & $5.3 \times 10^{-6}$ &  44.498 & 0.504 \\
		   & $(out,out)$ & -0.228 & 0.001 &-0.235 & $5.4 \times 10^{-6}$ &  44.105 & 0.499  \\
		   & $(in,in)$ & -0.227 & 0.001 & -0.234 & $4.6 \times 10^{-6}$ &  44.207 & 0.501\\   
		   \hline

Bipartite& $(out,in)$ & -0.974 & 0.001 & -0.715 & $4.7 \times 10^{-5}$ &  -59.537 & -0.511 \\
		   & $(in,out)$ & -0.973 & 0.001 & -0.705 & $9.6 \times 10^{-5}$ &  -56.944 & -0.488 \\
		   & $(out,out)$ & -0.974 & 0.001 &-0.711 & $9.6 \times 10^{-5}$ &  -58.222 & -0.499  \\
		   & $(in,in)$ & -0.973 & 0.001 & -0.710 & $5.3 \times 10^{-6}$ &  -58.514 & -0.502 \\   
		   \hline
\end{tabular}
\end{table*}

\begin{table*}
\caption{Standard deviations in food-web models.  We show the standard deviations in $r(\alpha,\beta)$ for $500$ instances per real-world network of the cascade and niche model.  Instances are constructed according to the procedure described in the \textit{Materials and Methods}, following ref. \cite{mar}; note the large standard deviations of the niche model. \newline}

\begin{tabular}{|c|c| | c|c|c|c|c|c|c|}
\hline
Network & $(\alpha,\beta)$ &$\sigma_r^{\rm{cascade}}$ &$\sigma_r^{\rm{niche}}$ \\
\hline \hline \hline
Coachella & $(out,in)$ & 0.0268 & 0.1501  \\
		   & $(in,out)$ & 0.0235 & 0.0826 \\
		   & $(out,out)$ & 0.0289 & 0.1033  \\
		   & $(in,in)$ & 0.0262 & 0.0739\\
		   \hline
Little Rock & $(out,in)$ & 0.0178 & 0.1314 \\
		   & $(in,out)$ & 0.0127 & 0.0354 \\
		   & $(out,out)$ & 0.0173 & 0.0777   \\
		   & $(in,in)$ & 0.0166 & 0.0642 \\
		   \hline
St. Marks & $(out,in)$ & 0.0583 & 0.1849  \\
		   & $(in,out)$ & 0.0455 & 0.0729  \\
		   & $(out,out)$ & 0.0592& 0.1341   \\
		   & $(in,in)$ & 0.0592 & 0.1046 \\  
\hline
St. Martin & $(out,in)$ & 0.0575 & 0.1841  \\
		   & $(in,out)$ & 0.0436 & 0.0759  \\
		   & $(out,out)$ & 0.0603 & 0.1276   \\
		   & $(in,in)$ & 0.0582 & 0.1038  \\  
\hline
Ythan & $(out,in)$ &0.0486 & 0.1636 \\
		   & $(in,out)$ & 0.0342 & 0.0566  \\
		   & $(out,out)$ & 0.0463 & 0.1116   \\
		   & $(in,in)$ &0.0467 & 0.0954  \\  
		   \hline
\end{tabular}
\end{table*}



\end{document}